# Smith-Purcell Radiation from Low-Energy Electrons

Aviram Massuda[1,2], Charles Roques-Carmes[1,2], Yujia Yang[2], Steven E. Kooi[3], Yi Yang[1,2], Chitraang Murdia[1], Karl K. Berggren[2], Ido Kaminer[1], Marin Soljačić[1]

[1]Department of Physics, Massachusetts Institute of Technology, 77 Massachusetts Avenue, Cambridge, Massachusetts 02139, USA

[2]Department of Electrical Engineering, Massachusetts Institute of Technology, 50 Vassar Street, Cambridge MA 02139, USA

[3]Institute for Soldier Nanotechnologies, 77 Massachusetts Avenue, Cambridge, Massachusetts 02139, USA

**Recent advances in the fabrication of nanostructures[1-3] and nanoscale features in metasurfaces[4-6] offer a new prospect for generating visible, light emission from low energy electrons. In this paper, we present the experimental observation of visible light emission from low-energy free electrons interacting with nanoscale periodic surfaces through the Smith-Purcell (SP) effect[7]. SP radiation is emitted when electrons pass in close proximity over a periodic structure, inducing collective charge motion or dipole excitations near the surface[7-9], thereby giving rise to electromagnetic radiation. We demonstrate a controlled emission of SP light from nanoscale gold gratings with periodicity as small as 50 nm, enabling the observation of visible SP radiation by low energy electrons (1.5 to 6 keV), an order of magnitude lower than previously reported[4, 7, 10-14]. We study the emission wavelength and intensity dependence on the grating pitch and electron energy, showing agreement between experiment and theory. Further reduction of structure periodicity[15] should enable the production of SP-based devices that operate with even slower electrons that allow an even smaller footprint and facilitate the investigation of quantum effects for light generation in**

**nanoscale devices[16]. A tunable light source integrated in an electron microscope would enable the development of novel electron-optical correlated spectroscopic techniques, with additional applications ranging from biological imaging to solid-state lighting[17].**

Tunable nanoscale light sources are of utmost importance for nanophotonics. Free-electron-driven light sources offer a promising avenue for achieving this goal[1, 12, 18, 19]. These devices benefit from flexible material choices as well as from the ability to focus electrons to nanoscale spots, which in turn enables the tailoring efficient of interactions with nanoscale structures. Thus far, most of the free-electron radiation sources have used relativistic electrons, from highly relativistic energies as in synchrotrons[20] and free electron lasers[21-24] to modestly relativistic energies as radiation sources in the microwave[25, 26] and visible[7, 8, 12]. The requirement for large electron velocities in those conventional setups has kept free-electron light sources away from compact or on-chip applications.

In this work, we leverage recent advances in nanoscale fabrication techniques that have enabled the study of new fundamental effects and their applications involving the interaction of free electrons with light and matter. For instance, new opportunities to explore the SP effect in nanoscale structures like plasmonic arrays[14, 27] and metasurfaces[28] have been recently investigated. Such systems can be used as sources of visible and IR light that are tunable by adjusting the electron velocity[7]. The possibility of observing *shorter* wavelength emissions from relatively low-energy electrons (accessible with regular scanning or transmission electron microscopes, SEM or TEM) is a very promising field of research[4, 18, 29-31], because of the exciting applications of short wavelength radiation in beam diagnostics[31], particle detection[32], biological imagining in the water window[33] and nanolithography[34].

SP radiation is emitted when an electron passes in close proximity over a periodic surface, inducing charges at the surface of the grating to rearrange themselves to screen the field of the moving electron, thereby inducing the emission of electromagnetic radiation[7-9]. In 1953[7] Smith and Purcell measured for the first time electromagnetic radiation produced by a free-electron beam passing over a metallic grating. They found that radiated wavelength depends on the properties of the structure and of the exciting electron beam and is given by the following well-known formula:

$$\lambda = \frac{a}{m}\left(\frac{1}{\beta} - \cos\theta\right) \tag{1}$$

where $\lambda$ is the radiation wavelength, $a$ is the grating pitch, $m$ is the diffraction order, $\beta = v/c$ is the normalized speed of electron passing over the structure, and $\theta$ is the angle of emission measured from the direction of beam propagation.

Smith and Purcell used relatively high-energy electrons with $\beta \sim 0.8$ for their original experiment. Further experimental demonstrations of the SP effect confirmed that it occurs over a wide spectral range. SP radiation has been demonstrated experimentally in the far infrared[35] and in the THz and mm-wave regimes[36, 37]. However, only a limited effort has been devoted to studying the SP radiation in the visible region (e.g., references 4, 7, 8). Up until now, these experimental demonstrations of SP radiation relied on the use of electrons that are moderately or highly relativistic with the lowest electron energy to generate SP radiation was 12 keV[27].

Relatively large-pitch periodic structures were used to record SP radiation, with the original experiment using metallic gratings of 1.67 μm pitch, to more recent literature scaling the structure periodicity down to 130 nm[14]. Reducing the periodicity to 50 nm enables us to observe optical SP radiation produced by nonrelativistic electron energies (about 5-10% the speed of light), and with an order of magnitude lower than the pervious record[27].

The key challenge to observing optical SP radiation from a nanograting is aligning the electron beam with a nanograting that is limited in size by nanofabrication constrains (typically about 200 μm along each lateral dimension parallel to the surface). We overcome this challenge by using a proprietary setup that we have developed. The setup makes it possible to spatially resolve the light emission simultaneously with collecting the emission spectra[27]. This way we can image the location of electron interaction with the surface and thus align the electron beam path with the nanograting and maximize the interaction between the two.

We have fabricated gold grating samples with 50 nm and 60 nm periods using electron beam lithography (EBL) and a lift-off technique. They consist of gold lines on a thick gold layer over a 200 μm × 200 μm area. The samples were mounted inside the chamber of a modified SEM experimental setup (Figure 1) and the electron kinetic energy was tuned between 1.5 and 6 keV.

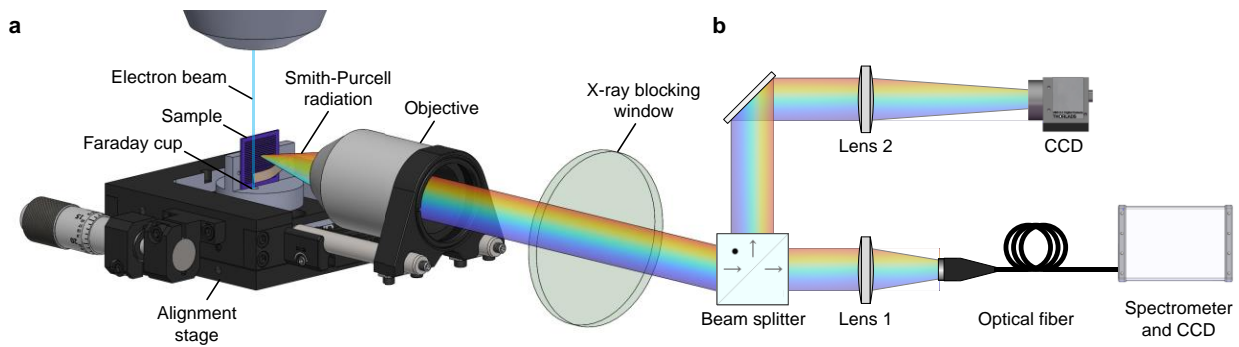

**Figure 1 Illustration of our SEM-based experimental setup used to observe SP radiation.** (a) Inside the SEM vacuum chamber the sample is held so that its surface is almost parallel to the path of the electron beam. The emitted light is collected by an objective and (b) directed to a beam splitter (BS) splitting the optical beam to an optical fiber collector that leads to a spectrometer (Lens 1) and to a CCD camera (Lens 2) that images the surface of the sample.

Figure 2 shows the measured SP spectra from a 50 nm pitch gold grating sample, with the peaks compared to the theoretical predication (equation (1)) down to electron energy of 1.5 keV. In addition to the tunable SP peaks, the spectra show a strong cathodoluminescence (CL) background around 550 nm.

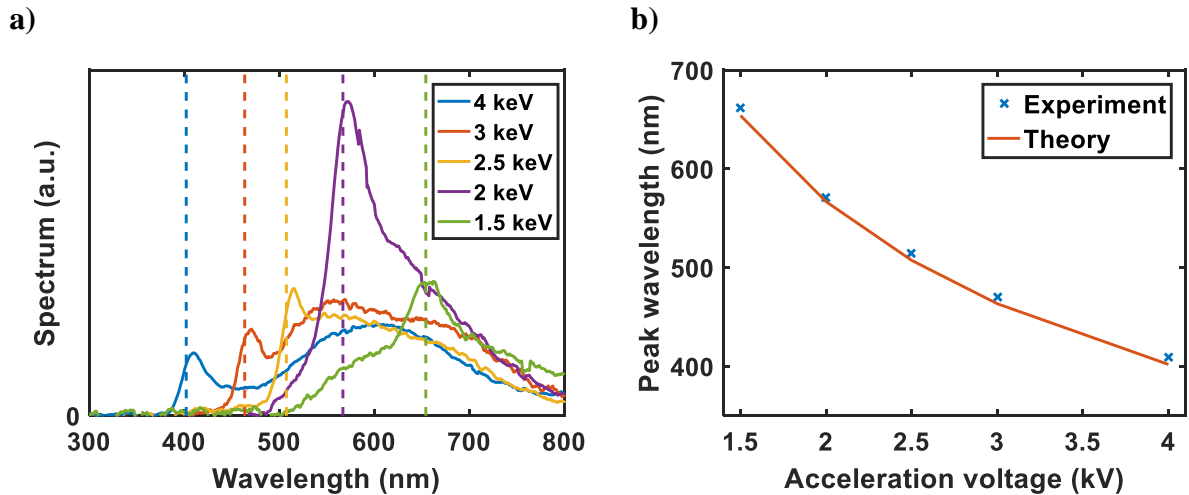

**Figure 2 SP radiation from low electron velocities and small pitch grating.** (a) Measured spectra for different kinetic energies from a grating with 50 nm pitch. The dashed vertical lines are calculated according to the conventional SP theory at normal emission [7], with colors corresponding to different kinetic energies. (b) Peak wavelength comparison between experiment and theory.

Figure 3b shows the measured SP spectra from 60 nm pitch gold nanogratings (shown in figure 3a), with the peaks compared to the theoretical prediction (equation (1)) at electron energies of 2-6 keV.

Using power calibration measurements (see methods section), we estimate power levels on the order of ~1 nW with a beam current of 100 nA (integrating the SP spectral power to obtain the total power)

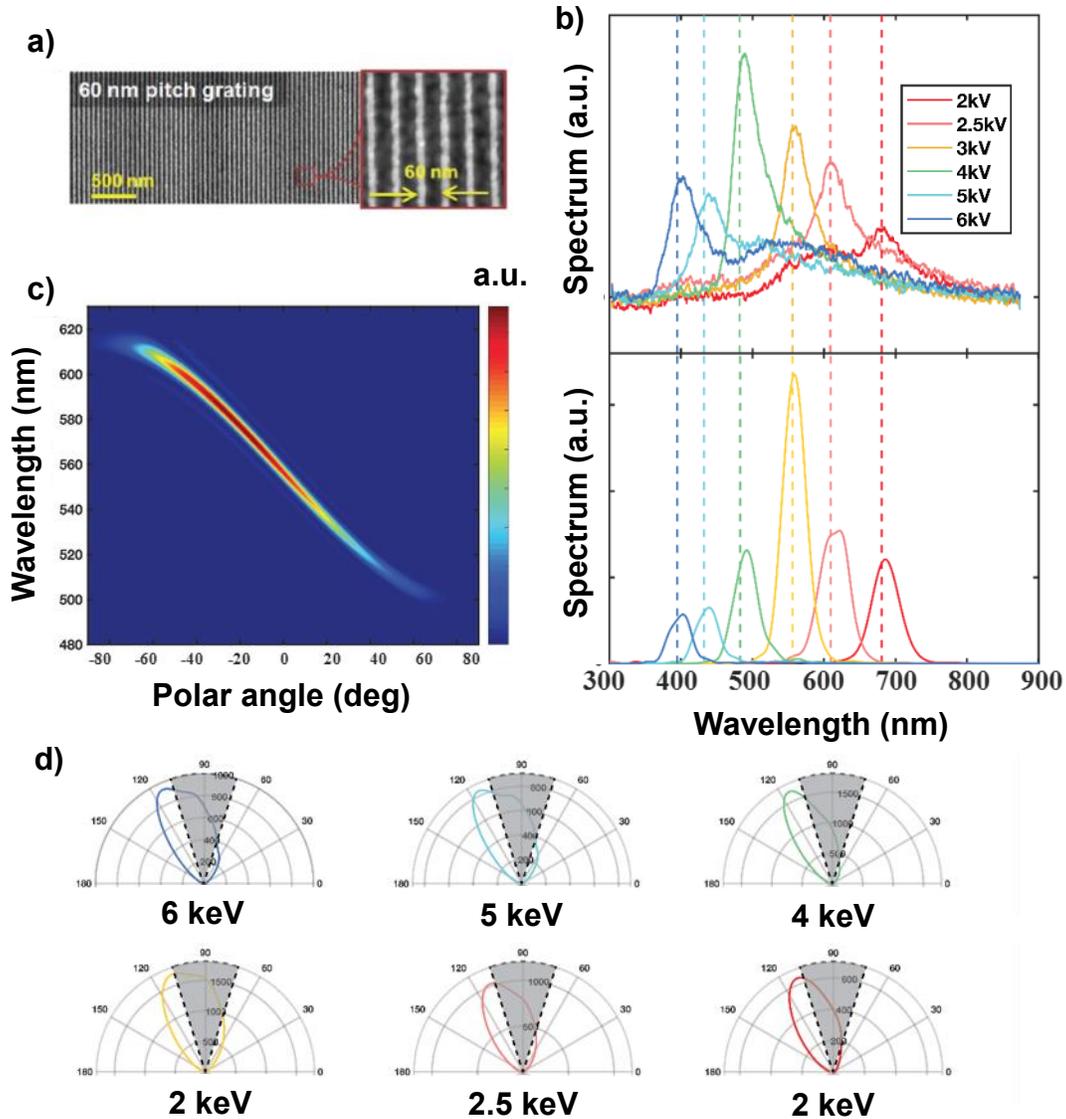

**Figure 3 SP radiation from nonrelativistic electrons: measured vs simulated radiation.** (a) SEM images of 60 nm pitch gratings (the red square shows a zoomed in display). (b) Upper plot: Measured spectra for different kinetic energies for a 60 nm pitch. Lower plot: Time-domain far-field distribution computed with $N=20$ unit cells (estimated number of unit cells with which each electron in the beam interacts) and integration over an angle corresponding to the numerical aperture of the objective used in our experiment (NA=0.3). The dashed vertical lines are calculated according to the conventional SP theory at normal emission [7], with the color corresponding to the same kinetic energies. (c) Simulation of the angular distribution for a kinetic energy of 3 keV and a grating of $N$=100 unit cells. The polar angle is measured from the direction normal to the beam propagation. At each angle, we verify the wavelength radiated with maximum intensity corresponds to the theoretical predication[7]. (d) Angular distribution of the SP emission of the same setup as in (c). The polar is angle is measured from the direction of propagation of the electron beam. We observe that the radiation for this grating pitch and energies is most efficient backward. The shaded grey area corresponds to the numerical aperture of the objective used in our experimental setup.

In figure 2 and figure 3b, gold gratings of period $a$=50 and 60 nm were shown to produce SP emission over the 395-654 nm wavelength range with low energy-electrons (1.5-6 keV). To the best of our knowledge, these periodicities and electron kinetic energies used to generate SP radiation are the smallest reported so far (by a factor of 2.6 in the pitch and 8 in the kinetic energy). Our experimental observations are confirmed by simulations (figure 3c and figure 3d), which also predict the spectrum lineshape and take into account the structure geometry and the optical response of the material. The simulation computes the scattering spectrum of the evanescent field carried by the electron and verifies the typical cosine-like shape of the original formula (equation (1)) as shown in figure 3c. Interestingly, due to the optical response of the nanograting at the optical frequencies, the most efficient wavelengths of emission are usually emitted at some backward angle ($\theta > \frac{\pi}{2}$) (see figure 3d). This fact explains the slight red-shift between the theoretical prediction at normal emission (dashed lines) and the peak wavelength, as can be seen in both our experimental and simulation results (figure 3b).

An advantage of using slow electrons is their higher photon extraction efficiency $\eta = \frac{dN/d\omega}{E_k}$, where $dN/d\omega$ is the radiation intensity (generated number of photons per frequency) and $E_k$ is the kinetic energy of electrons. To exemplify this advantage, we compute these quantities for a fixed radiation wavelength of 700 nm (figure 4a). The electron structure separation is taken to be the same as the pitch of the nanograting and the radiation is calculated for one unit cell, so that the geometry is invariant by scaling with $\beta$. The results of this calculation shows that the radiation intensity decreases with smaller electron velocities, so slow electrons radiate less than fast electrons (blue curve in figure 4a). On the other hand, the extraction efficiency increases with smaller electron velocities, so slow electrons radiate more efficiently than fast electrons (orange curve in figure 4a); for example, the extraction efficiency for $\beta = 0.1$ is one order of magnitude higher than the extraction efficiency for $\beta = 0.6$.

Finally, we experimentally observe the SP spectral peak at each fixed emission wavelength gets narrower for smaller electron velocities. i.e., when fixing the emission wavelength, lower electron velocities give a narrower emission spectrum. Figure 4b shows this effect by showing the SP radiation at a fixed wavelength (700 nm) from three different nanograting pitches with three different electron velocities. This is consistent with the SP formula (equation (1)), as it implies that the spectral bandwidth of SP radiation is linear with the electron velocity. This can be seen by fixing the peak wavelength $\lambda_{\text{peak}} = \frac{a}{m}\frac{1}{\beta}$ and the collection numerical aperture, and then noticing that equation (1) gives $\Delta\lambda \propto \frac{a}{m} = \beta\lambda_{\text{peak}}$. The effect of SP bandwidth narrowing for slow electrons also bolsters the interest of observing SP radiation from even lower energy non-relativistic electrons, yielding an additional degree of freedom to shape the spectral response of SP radiation.

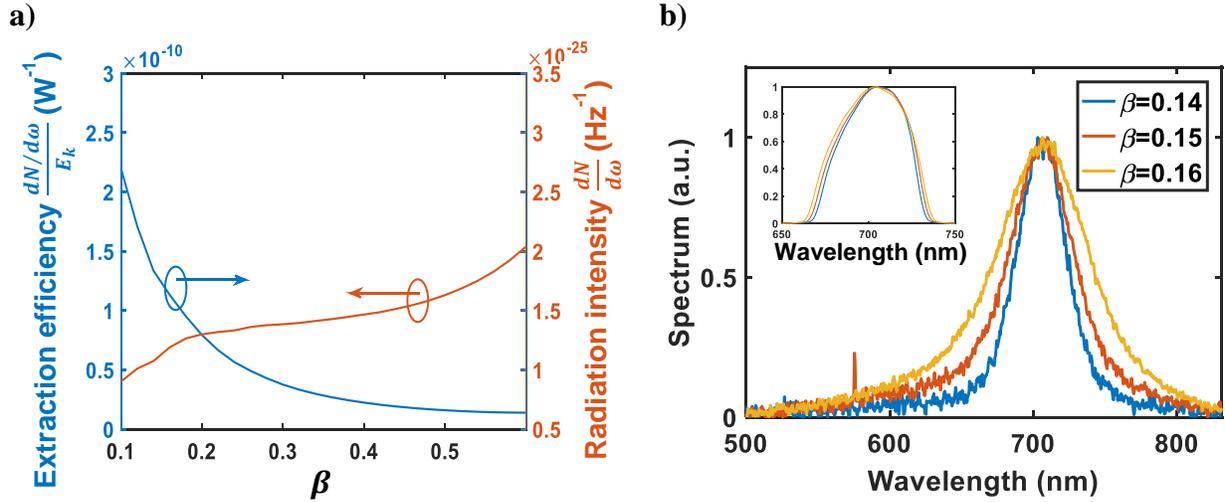

**Figure 4 Characterization of SP radiation for slow electrons.** a) Simulated efficiency as a function of $\beta$ with $\lambda = L/\beta$ fixed at 700 nm. The blue curve represents the power spectrum, i.e. the total number of photons/Hz/period/electron measured at a fixed wavelength of 700 nm. The red curve represents the relative efficiency of the SP emission defined as the ratio of the emitted photon energy to the kinetic energy of the incident electron. b) Increase of spectral coherence (bandwidth reduction): the normalized radiated spectrum for different $\beta$.

In conclusion, while most of the literature has focused on moderately or highly relativistic electrons, the prospect of achieving a low-electron-energy nanoscale light source paves the way for new regimes of light-matter interaction. Utilizing a modified SEM that enables resolving the spatial and spectral information of the light emission simultaneously, we observe SP from nanoscale gratings, which enabled us to drastically reduce the electron beam energies, pushing towards the development of efficient on-chip tunable light sources. With the ability to fabricate ever-reducing feature sizes, our work provides a platform to bring SP radiation from accelerator physics and high-energy electron physics to integrated devices. A similar motivation has recently led to reducing the velocity threshold of Cherenkov radiation[29]. The unique prospect opened up by compact new SP sources of light lie in their tunability since their emission wavelength can be controlled by the electron velocity, and can reach spectral ranges that conventional light sources cannot commonly achieve. These include the EUV and soft X-ray radiation ranges, with numerous exciting applications[23]. So far, SP radiation has yet to be experimentally demonstrated in these regimes, even though there are promising predictions towards the realization of an efficient SP source in the UV regime [38, 39].

## **Methods**

**Fabrication.** SP gratings were fabricated on Au coated Si substrates. An Au coating layer was deposited by electron-beam evaporation of 5 nm of Ti and then 200 nm of Au onto a Si chip. A ~70 nm film of PMMA was spin-coated onto the Au coated Si chip and then soft-baked at 180 °C. Grating patterns were produced by an Elionix F-125 electron beam lithography system using an accelerating voltage of 125 keV beam current of 500 pA. Exposed PMMA was developed in 3:1 IPA:MIBK at 0 °C for 30 s (See reference 40) and then dried with flowing $N_2$ gas. 20 nm of Au was then deposited via electron-beam evaporation. Metal lift-off was performed in NMP at 50 °C for approximately 60 min during which the sample was gently rinsed with flowing NMP. After lift-off, the sample was rinsed with acetone and IPA. Finally, gentle O2 plasma ashing (50 W, 60 s) was applied to remove residual resist and solvents.

**Experimental setup.** The experimental setup used for this experiment is shown in Figure 1. We used a JEOL JSM-6010LA scanning electron microscope (SEM) that we modified to enable the alignment of free electrons to pass in close proximity to the surface of periodic structures and the emitted light was collected with a Nikon TU Plan Fluor 10x microscope objective with a numerical aperture (NA) of 0.30. The emitted SP photons exited the SEM chamber via a leaded glass window and were detected with a spectrometer (Acton SP-2360) equipped with a low noise thermoelectrically cooled CCD (Princeton Instrument PIXIS-400B). The spectra were collected using a grating with a line density of 150 g/mm and a blaze angle of 500 nm while using a low noise ADC at the rate of 100 kHz. The exposure time was 600 ms and the signals were averaged over 20 repetitions. The beam currents were measured using Keithley 6485 picoammeter connected to a Faraday Cup mounted on a SEM mount.

**Calibration.** In order to estimate the absolute emitted optical power of the SP radiation we performed a power calibration measurement. Using the same experimental configuration of the SP measurements, a calibrated source (AvaLight-HAL-CAL) for the visible range (350-1095 nm) was placed at the same location as the sample. The signal measure for the calibrated light source was obtained at the spectrometer in units of signal counts. The power calibration profile was obtained by measuring the calibrated source using optical spectrum analyzer (AQ-6315A). The experimental setup response function was obtained by dividing the measured profile of the calibration source by its calibration profile. The SP spectral power from a sample was then obtained by dividing the measured signal from the sample by the setup response function.

**Time and frequency domain representation of the electron beam.** The electron beam can be represented as a time-dependent propagating point-electron: $J(\vec{r}, t) = -e v \delta(x - vt)\delta(y - y_0)\delta(z - z_0) \vec{x}$. Taking the Fourier transform of this time-dependent distribution, we get the following frequency-domain representation and the associated polarization distribution: $J(\vec{r}, \omega) = -e \exp(-i\frac{\omega x}{v}) \delta(y - y_0)\delta(z - z_0) \vec{x}$ and $(\vec{r}, \omega) = i\frac{e}{\omega}\exp(-i\frac{\omega x}{v}) \delta(y - y_0)\delta(z - z_0) \vec{x}$ . In time-domain simulations, we can use a set of closely spaced dipole sources in order to mimic the propagation of the electron beam, and compare these results with frequency-domain simulations where we can directly implement a line current.

**Time-domain simulations.** This correspondence allows us to represent an electron beam as a set of closely spaced dipoles shifted in time. Time-domain simulations are run using the commercial FDTD software Lumerical, where a dipole in frequency domain is defined by its source norm $s(\omega)$ and base amplitude $p_{base}$ as $p_k(\omega, \vec{r}) = p_{base} s(\omega) \exp(-\frac{i\omega x_k}{v})$, where $x_k$ is the position of the $k$−th dipole and the exponential term in frequency-domain corresponds to a time-delay in time-domain. The induced polarization by the set of dipoles will be $\sum_{i=1}^{N_{dip}} p_k(\omega, r) \xrightarrow{N_{dip} \to +\infty} P(\vec{r}, \omega)$. To ensure the field recorded is generated by the emission of one electron of charge $e$, we normalize the recorded electric field by $\alpha$ and power by $\alpha^2$ where $\alpha = \frac{e \Delta x}{p_{base} s(\omega) \omega}$. We usually measure the power on a plane and get a result $P(\omega)$ in units of $\frac{W}{Hz^2}/m^2$. Integrating this result on a surface already simplifies the units to $W/Hz^2$. This result can be converted in number of photons (per electron): $N_{SP} = \int \frac{P(\omega)}{\hbar\omega} d\omega$.

In Figure 3b, simulation parameters are designed to match the experimental setup. For a mean tilt angle of the electron beam of 1°, a pitch of 60 nm, and assuming the electron beam effectively interacts with the grating at a distance H < 5 nm (since the interaction efficiency drops exponentially with the pitch as $e^{-4\pi H/a}$), we get an effective amount of unit cells of N=20. The field is recorded 1 μm above the grating and projected in the farfield. Only radiation emitted at an angle less than the numerical aperture of the objective used in the experiment contributes to the spectrum plotted in figure 3b. In Figure 3d, a simulation setup similar to figure 3b is used to measure the angular pattern of the radiation.

**Frequency domain simulations.** Figure 4b is calculated using finite-element method (COMSOL Multiphysics). Electrons are treated as a line current (see the Fourier transform in the time-domain simulation) with periodic boundary condition imposed.

## Acknowledgements

We thank Dr. Richard Hobbs, Dr. Philip Keathley, Dr. William Putnam and Dr. Chung-Soo Kim for useful discussion. We also thank Mark Mondol and Jim Daley for assistance with fabrication and Dr. Peter Krogen for assistance with the experimental setup. The work was supported by the Army Research Office through the Institute for Soldier Nanotechnologies under contract No. W911NF-13-D-0001. Yi Yang was partly supported by the MRSEC program of the National Science Foundation under grant no. DMR-1419807. The research of Ido Kaminer was partially supported by the Seventh Framework Programme of the European Research Council (EP7-Marie Curie IOF) under Grant No. 328853-MC-BSiCS. Karl K. Berggren and Yujia Yang would like to acknowledge generous support of this research by Arthur Chu and Jariya Wanapun.